%% file: comonads.tex
\pdfoutput=1

\documentclass[envcountsame]{llncs}

\usepackage[utf8]{inputenc}

\usepackage[protrusion=true,expansion=true]{microtype}

\newcommand{\parencite}[1]{\cite{#1}}
\newcommand{\textcite}[1]{\cite{#1}}

\usepackage{ownstylesansthm}

\usepackage{comment}
\includecomment{Long}\excludecomment{Short}

\begin{document}

\title{Terminal semantics for codata types\\in intensional Martin-L\"of type theory}

\author{Benedikt Ahrens and R\'egis Spadotti}

\institute{
Institut de Recherche en Informatique de Toulouse\\
Universit\'e Paul Sabatier, 
Toulouse}

\newcommand{\fat}[1]{\textbf{#1}}

\maketitle


\begin{abstract}

 In this work, we study the notions of \emph{relative comonad} and \emph{comodule over a relative comonad}, and  
 use these notions to give a terminal coalgebra semantics for the coinductive type families of streams and
 of infinite triangular matrices, respectively, in intensional Martin-L\"of type theory.
 Our results are mechanized in the proof assistant \coq.

  \end{abstract}

\section{Introduction}
 
 In this work, we study the notions of \emph{relative comonad} and \emph{comodule over a relative comonad}.
 We then use these notions to characterize several \emph{co}inductive data types in intensional Martin-L\"of type theory
 via a universal property.
 
 In a set-theoretic setting, inductive sets are characterized as initial algebras for 
 some endofunctor on the category of sets. For instance, the set of natural numbers constitutes the carrier of the 
 initial algebra of the functor $X \mapsto 1 + X$.
 
 In a type-theoretic setting as given by Martin-L\"of type theory \parencite{martin_lof}, 
 two approaches to the semantics of inductive types have been studied: 
 one approach consists in showing that inductive types exist in a \emph{model} of the type theory,
 as is done by \textcite{DBLP:journals/apal/MoerdijkP00}.
 Another approach is to prove that adding certain type-theoretic rules to the type theory
 implies (or is equivalent to) the existence of a universal object \emph{within} type theory
 (see, e.g., \parencite{DBLP:conf/lics/AwodeyGS12,DBLP:journals/tcs/Dybjer97}).
 This latter approach is the one we take in the present work.
 
 Some attention has to be given to 
 the precise formulation of the type theory in question:
 One important feature of Martin-L\"of type theory is the \emph{identity type}, a type family that associates to any two 
 inhabitants $a,b : A$ of a same type $A$ the type of \enquote{identities} between them.
 One distinguishes \emph{extensional} and \emph{intensional} type theory, according to whether terms of identity type
 are reflected into the internal, judgmental equality of the type theory or not. This difference must be considered when defining the notion of
 \enquote{initial algebra}, in particular the \emph{uniqueness} part of initiality:
 
 In extensional type theory, propositional equality as given by the Martin-L\"of identity type \parencite{martin_lof} is reflected 
 into judgmental equality via a \emph{reflection rule}.
 This reflection rule equips \emph{extensional} type theory with extensional features similar to those of set theory.
 As a consequence, the characterization of a \textsf{W}-type---a member of a particular class of inductive types---in extensional MLTT as initial algebra for some endofunctor on the 
 category of types \parencite{DBLP:journals/tcs/Dybjer97} works as in the category of sets.
 Indeed, in extensional MLTT, one has function extensionality available, which suffices to 
 deduce that there is a \emph{judgmentally unique} algebra morphism from the initial algebra to any algebra.

 Intensional Martin-L\"of type theory \parencite{martin_lof} lacks this reflection principle for the sake of decidability of type checking. 
 It forms the base of two computer proof assistants, \coq and \agda.
 \textsf{W}-types have been studied \parencite{DBLP:conf/lics/AwodeyGS12} in \emph{Homotopy Type Theory} (HoTT) \parencite{hottbook}, an extension of intensional Martin-L\"of type theory. 
 In this extension, function extensionality is provable from the \emph{Univalence Axiom}. 
 For a suitable definition of \emph{uniqueness}---\emph{contractibility} in HoTT jargon---one can then prove uniqueness of the
 algebra morphisms out of the one whose carrier is given by the \textsf{W}-type.
 The mentioned work \parencite{DBLP:conf/lics/AwodeyGS12} thus shows that the characterization of \textsf{W}-types as initial algebras carries over 
 from extensional to intensional type theory if one adds an extensionality principle for functions and adapts the notion of uniqueness.

 The characterization of inductive sets/types as initial objects in some category
 has been extended to some \emph{heterogeneous}---also called \emph{nested}---inductive data types, e.g., the type of $\lambda$-terms,
 in different formulations \parencite{fpt,DBLP:journals/iandc/HirschowitzM10}.
 The main goal of these works is not just to characterize a data type via a universal property, but rather a data type
 \emph{equipped with a canonical, well-behaved substitution operation}.

 Dually to inductive sets, \fat{co}inductive sets such as streams are characterized as terminal objects \parencite{jacobs1997tutorial}.
 Inhabitants of such sets are equal if and only if they are \emph{bisimilar} \parencite{DBLP:journals/mscs/TuriR98}:
 Intuitively, two elements of a coinductive set are the same if they allow for the same observations.
 
 This correspondence between equality and bisimilarity fails in IMLTT, when equality is considered to be given by the 
 Martin-L\"of identity type. Instead, one defines \emph{bisimilarity} as a coinductive predicate 
 on a coinductive type, and one reasons about the terms of a coinductive type modulo the bisimilarity predicate
 rather than identity \parencite{DBLP:conf/types/Coquand93}.
 Consequently, we consider two maps into a coinductive type to be the same if they are \emph{pointwise bisimilar}---an analogue
 to the aforementioned principle of function extensionality. 
 With these conventions, we give, in the present work, a characterization of some coinductive data types as \emph{terminal} objects in some category 
 defined in intensional Martin-L\"of type theory.
 More precisely, we consider an example of \emph{homogeneous} codata type, streams, and 
 an example of \emph{heterogeneous} codata type, triangular matrices.
 For each of these examples we prove, 
 from type-theoretic rules specifying the respective codata type added to the basic rules of Martin-L\"of type theory,
 the existence of a terminal object in some category \emph{within IMLTT}.
 Our terminal semantics characterizes not only the codata types themselves but also the bisimilarity relation and
 a canonical cosubstitution operation on them.
 
 The fact that cosubstitution for coinductive data types is comonadic in a set-theoretic setting is established by \textcite{DBLP:conf/sfp/UustaluV01}.
 In IMLTT however, in order to characterize that cosubstitution operation on a given codata type, and its algebraic properties,
 we develop the notion of \emph{relative comonad} and \emph{comodule over a relative comonad}.
 The need to consider \emph{relative} comonads arises from the  need to check the algebraic properties of cosubstitution modulo \emph{bisimilarity} rather
 than modulo identity (in the sense of ML identity types).

 All our results have been implemented in the proof assistant \coq \parencite{coq84pl3}.
 The \coq source files and HTML documentation are available online \parencite{trimat_coq}.
 In this document, we hence omit the proofs and focus on definitions and statements of lemmas.
\begin{Short}
 A longer version of this article is available on the arXiv \parencite{trimat_coq}.
\end{Short}

 \subsubsection*{Disclaimer}The category-theoretic concepts studied in this work are agnostic to the foundational system being worked in.
 While we present them in a type-theoretic style, the definitions and lemmas can trivially be transferred to a set-theoretic setting.
 Throughout this article, we use type-theoretic notation,  writing $t:T$ to indicate that $t$ is of type $T$. 
 For instance, we write $f : \C(A,B)$ to indicate that $f$ is a morphism from object $A$ to object $B$ in category $\C$.
 Whenever an operation takes several arguments, we write some of them as indices; these indices might be omitted when 
 they can be deduced from the type of the later arguments.
 We assume basic knowledge of category theory; any instances used are defined in the following.
  
 \subsubsection*{More related work}
 The notion of \emph{module over a monad}, which we dualize and generalize in this work, is used by \textcite{DBLP:journals/iandc/HirschowitzM10}
 to give an initial semantics result for languages with variable binding. 
 Their work is based on work of \textcite{alt_reus},
 who show that the lambda calculus equipped with a simultaneous substitution constitutes a monad.
 We make use of the notion of \emph{relative comonad}, the dual to relative monads as introduced by \textcite{DBLP:conf/fossacs/AltenkirchCU10}.
 One of our main examples, the codata type of infinite triangular matrices, is studied by \textcite{DBLP:conf/types/MatthesP11}.
 Redecoration for both finite and infinite triangular matrices is used by \textcite{DBLP:journals/tcs/AbelMU05} to exemplify 
 the expressivity of the studied recursion schemes.
 
\begin{Long}
 
 \subsubsection*{Organisation of the paper}
  In \Cref{sec:preliminaries} we introduce some concepts and notations used later on.
  In \Cref{sec:tri} we present the coinductive type families $\stream$ of streams and $\Tri$ of infinite triangular matrices and some operations on those codata types.
   Their specifying rules are given in \Cref{stream_rules} and \Cref{tri_rules}, respectively.
  In \Cref{sec:comonads} we present \emph{relative comonads} and define the category of comonads relative to a fixed functor.
    We give some examples of such structures, using the codata types presented in \Cref{sec:tri}.
  In \Cref{sec:comodules} we define comodules over relative comonads and give some constructions of comodules.
     Again, examples of such structures are taken from \Cref{sec:tri}.
  In \Cref{sec:coalgebras_for_tri} we define categories of coalgebras for the codata types presented in \Cref{sec:tri},
      based on the category-theoretic notions developed in the previous sections.
      We then prove that the codata types constitute the terminal coalgebras in the respective categories.
      Finally, we present an example of a map defined as a terminal map exploiting the universal property of streams.
  In \Cref{sec:formal} we explain some details of the formalization of this work in the proof assistant \coq.
  A table with the correspondence between formal and informal definitions is given in \Cref{sec:table_formal_informal}.

\end{Long}

\section{Preliminaries}\label{sec:preliminaries}

In this section we present some particular categories and functors used later on, and fix some notation.

\begin{definition}[Some categories]\label{def:set_setoid}
 We denote by $\Set$ the category of types (of a fixed universe) and total functions between them in Martin-L\"of type theory. 
 A morphism $f$ in this category is denoted by $f : A \to B$.
 
 We denote by $\Setoid$ the category an object of which is a \emph{setoid}, i.e.\ a type equipped with an equivalence relation.
 A morphism between setoids is a type-theoretic function between the underlying types that is compatible in the obvious sense with the equivalence relations of the source and target setoids.
 If $A$ is a setoid, we also use $A$ to refer to its underlying type, and thus write $a:A$ for an element $a$ of the type underlying the setoid $A$. 
 We write $a\sim a'$ for related elements $a$ and $a'$ in $A$.
 We consider two parallel morphisms of setoids $f,g:A\to B$ equal if for any $a:A$ we have $fa \sim ga$.
 
 We also write $f:A\to B$ for a morphism $f$ between objects $A$ and $B$ in some category, in particular in the category of types.
 \end{definition}

\begin{definition}\label{def:eq}
 The functor $\eq : \Set\to\Setoid$ is defined as the left adjoint to the forgetful functor $U : \Setoid \to \Set$.
  Explicitly, the functor $\eq$ sends any type $X$ to the setoid $(X,=_X)$ given by the type $X$ itself, equipped
  with the propositional equality relation $=_X$ specified via Martin-L\"of's identity type on $X$.
\end{definition}

\begin{remark}[Notation for product]
  We denote the category-theoretic binary product of objects $A$ and $B$ of a category $\C$ by $A\times B$.
  We write $\pr_1(A,B) : \C(A\times B, A)$ and $\pr_2(A,B) :\C(A\times B, B)$ for the projections, occasionally omitting the 
  argument $(A,B)$.
  Given $f : \C(A, B)$ and $g : \C(A,C)$, we write $\langle f,g\rangle : \C(A,B\times C)$ for the induced map into the product such that
  $\comp{\langle f,g\rangle}{\pr_1} = f$ and $\comp{\langle f,g\rangle}{\pr_2} = g$.
\end{remark}

Both of the categories of \Cref{def:set_setoid} have binary products; they are \emph{cartesian monoidal}, i.e.\ the terminal 
object is neutral with respect to the product. Functors preserving the monoidal structure up to isomorphism
are called \emph{strong monoidal}:

\begin{definition}\label{def:monoidal_functor}
 A functor $F:\C\to\D$ between cartesian monoidal categories is \fat{strong monoidal} if, for any two objects $A$ and $B$ of $\C$,
  the morphism
 \[ \phi^F_{A,B} := \bigl\langle F(\pr_1), F(\pr_2) \bigr\rangle : \D\bigl(F(A\times B), FA\times FB\bigr)\enspace  \] 
 is an isomorphism.
 (Note that for \emph{cartesian} monoidal categories, the family $\phi$ of morphisms automatically 
  is compatible with the unitators and associators of the source and target categories, 
  since it is given by a universal property.)
\end{definition}

\begin{example}
  The functor $\eq: \Set \to \Setoid$ of \Cref{def:eq} is strong monoidal.
\end{example}

\section{Codata types in intensional Martin-L\"of type theory}\label{sec:tri}

We consider two particular coinductive type families in Intensional Martin-L\"of type theory (IMLTT) \parencite{martin_lof}, 
a type-theoretic foundational system.
For $a,b : A$, we denote by $a = b$ the Martin-L\"of identity type between $a$ and $b$.

In this section, we present these types, and we also define \emph{bisimilarity} for each codata type.
Bisimilarity is a coinductively defined equivalence relation on types which is considered 
as the appropriate notion of sameness on inhabitants of these types \parencite{DBLP:conf/types/Coquand93,DBLP:journals/corr/abs-cs-0603119}.
A coinductive type with bisimilarity hence forms a setoid as in \Cref{def:set_setoid}.
We thus denote bisimilar elements using an infix $\sim$, as in $t \sim t'$. 

Maps into a coinductive data type are specified by the observations, i.e.\ the value of the destructors, on the output of those maps.  
The precise rule for specifying maps into the considered coinductive type is given in the respective appendix.
In this text, we use a more convenient syntax, as illustrated in \Cref{eq:tail_sredec}.

The first example is the type of \emph{streams} of elements of a given base type $A$. 
The precise set of rules specifying that type is given in \Cref{stream_rules}.
In the presentation we use the notational convention of \Cref{def:set_setoid}, using the same name for a setoid and its underlying type.
\begin{example}\label{ex_stream}
  Let $A$ be a type. The type $\stream A$ of \emph{streams over $A$} is coinductively defined via the destructors 
  given in \Cref{fig:stream_destructors}.
  \begin{figure}[bt]
  \centering

     \def\extraVskip{3pt}
     \def\proofSkipAmount{\vskip.8ex plus.8ex minus.4ex}
    \AxiomC{$t : \stream A$} 
     \UnaryInfC{$\shead_A~t : A$}
      \DisplayProof
                        \hspace{3ex}
                                       \AxiomC{$t : \stream A$}
                                       \UnaryInfC{$\stail_A~t : \stream A$}
                                       \DisplayProof%
%
%
%
\hspace{3ex}
 \centering
                                            \def\extraVskip{3pt}
     \def\proofSkipAmount{\vskip.8ex plus.8ex minus.4ex}
    \AxiomC{$t \sim t'$} 
     \UnaryInfC{$\shead~t = \shead~t'$}
      \DisplayProof
                        \hspace{3ex}
                                       \AxiomC{$t \sim t'$} 
                                       \UnaryInfC{$ \stail~t \sim \stail~t'$}
                                       \DisplayProof   
  \caption{Destructors and bisimilarity for the coinductive family $\stream$} \label{fig:stream_destructors}
\end{figure}

   We define a \emph{co}substition operation $\sredec_{A,B} : (\stream A \to B) \to \stream A\to\stream B$ on streams via the following clauses:
   \begin{align} \comp{\sredec~f}{\shead} := f \quad\text{ and } \quad
                  \comp{\sredec~f}{\stail} := \comp{\stail}{\sredec~f} \enspace . \label{eq:tail_sredec}
    \end{align}
  We call such an operation \enquote{cosubstitution} since its type is dual to, e.g., the simultaneous substitution operation 
  of the lambda calculus \parencite{alt_reus}.
\end{example}

\begin{Long}

Streams are node-labeled trees where every node has exactly one subtree.
We also consider a type of trees where every node has an arbitrary, but fixed, number of subtrees, 
parametrized by a type $B$.

\begin{example}[Node-labeled trees]\label{ex_trees}
 We denote by $\Tree_B(A)$ the codata type given by one destructor $\shead$ and a family of 
 destructors $(\stail_b)_{b:B}$ with types analogous to those defining $\stream$ of \Cref{ex_stream}.
 We thus obtain $\stream$ by considering, for $B$, the singleton type.
\end{example}

\end{Long}

Another codata type we consider models
\emph{infinite triangular matrices}. It is more sophisticated than the type of streams as one of its destructors is \emph{heterogeneous}:

\begin{example}\label{ex_tri}
This codata type is studied in detail by
 \textcite{DBLP:conf/types/MatthesP11}.
 We give a brief summary, but urge the reader to consult the given reference 
 for an in-depth explanation. 
 The codata type family $\Tri$ of infinite triangular matrices 
 is parametrized by a fixed type $E$ for entries not on the diagonal, 
 and indexed by another, \emph{variable}, type $A$ for entries on 
 the diagonal. 
\begin{Long}
 Schematically, such a matrix looks like in \Cref{fig_tri}.
 \begin{figure}[bt]
 \centering
 \begin{tikzpicture}[scale = 0.6]
    \foreach \y in {0,...,2}
    {\foreach \x in {\y,...,2}
      \draw (\x+1, -\y) node[color=blue]{$E$} ;
    }
    \foreach \x in {-1,...,3} \draw (\x, -\x-1) node[color=red]{$A$} ;
    \foreach \x in {0,...,3} \draw (\x, -\x)
    node[color=blue]{\textbf{$E$}} ;
    \draw(4,-2) node{$\ldots$};

      \draw[color=purple!70]  (2.4, -4.3) --node[auto, swap, left]{$\cut$}
     (-1.2,-0.6) -- (2.8,-0.6);
    \draw[color=purple!70, dashed]  (2.8,-0.6) -- (3.6,-0.6);
    \draw[color=purple!70, dashed]  (2.4,-4.3) -- (3.0,-5);

    \draw (-2,0) node{$\head$} ;

    \draw[color=green] (2.8,0.4) -- (-0.5,0.4) -- (-0.5, -1.1) --node[auto, swap, left]{$\tail$}
    (2.8,-4.5)  ; 
    \draw[color=green, dashed] (2.8,0.4) -- (3.6,0.4) ; 
    \draw[color=green, dashed] (2.8,-4.5) -- (3.5,-5.3) ;

    \draw[color=orange](0, -0.5) ellipse (12pt and 25pt) ;  
    \draw[color=orange](1, -1.5) ellipse (12pt and 25pt) ;  
    \draw[color=orange](2, -2.5) ellipse (12pt and 25pt) ;  
    \draw[color=orange](3, -3.5) ellipse (12pt and 25pt) ;  
  \end{tikzpicture}\\[-2ex]
  \caption{An infinite triangular matrix over type $A$ and various operations}\label{fig_tri}
 \end{figure}
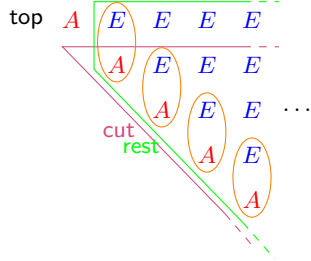
\end{Long}
  
 It is specified via two destructors $\head$ and $\tail$, whose types are given in \Cref{fig:tri_destructors}.
\begin{Long}
 Given a matrix over type $A$, its $\tail$---obtained by removing the first element on the diagonal, i.e.\ the $\head$ element---can 
 be considered as a trapezium as indicated by the green line in \Cref{fig_tri}, or alternatively, as
 a triangular matrix over type $E\times A$, by bundling the entries of the diagonal with those above as indicated by the orange frames in \Cref{fig_tri}.
 The latter representation is reflected in the type of the destructor $\tail$.
\end{Long}
\begin{Short}
 Given a matrix over type $A$, its $\tail$---obtained by removing the first element on the diagonal, i.e.\ the $\head$ element---can 
 be considered as a trapezium or, alternatively, as a triangular matrix over type $E\times A$, 
 by bundling the entries of the diagonal with those directly above the diagonal.
\end{Short}

 Bisimilarity on the inhabitants of that type is defined via the destructors of \Cref{fig:tri_destructors}.    
 As with streams, we denote by $\Tri A$ not only the resulting \emph{setoid} of triangular matrices over $A$, but also its
 underlying \emph{type}. 

\begin{figure}[bt]
  \centering

     \def\extraVskip{3pt}
     \def\proofSkipAmount{\vskip.8ex plus.8ex minus.4ex}
    \AxiomC{$t : \Tri~A$} 
     \UnaryInfC{$\head_A~t : A$}
      \DisplayProof
                        \hspace{3ex}
                                       \AxiomC{$t : \Tri~A$}
                                       \UnaryInfC{$\tail_A~t : \Tri(E\times A)$}
                                       \DisplayProof%
%
%
%
 \hspace{3ex}
                                            \def\extraVskip{3pt}
     \def\proofSkipAmount{\vskip.8ex plus.8ex minus.4ex}
    \AxiomC{$t \sim t'$}
     \UnaryInfC{$\head~t = \head~t'$}
      \DisplayProof
                        \hspace{3ex}
                                       \AxiomC{$t \sim t'$}
                                       \UnaryInfC{$ \tail~t \sim \tail~t'$}
                                       \DisplayProof   

  \caption{Destructors and bisimilarity for the coinductive family $\Tri$} \label{fig:tri_destructors}
\end{figure}
  A cosubstitution operation, \enquote{redecoration},
    $ \redec_{A,B} : (\Tri A \to B) \to \Tri A \to \Tri B$
  is defined  through the clauses
  \begin{align} \comp{\redec~f}{\head} := f \quad\text{ and } \quad
                  \comp{\redec~f}{\tail} := \comp{\tail}{\redec~(\extend~f)} \enspace . \label{eq:rest_redec}
    \end{align}
Here, the family of functions 
     $\extend_{A,B} : (\Tri A \to B) \to \Tri (E \times A) \to E\times B $
  is suitably defined to account for the change of the type of the argument of $\redec$ when redecorating $\tail~t : \Tri(E\times A)$
  rather than $t : \Tri A$, namely
  \[ \extend(f) := \langle \comp{\head_{E\times A}}{\pr_1(E,A)} , \comp{\cut_A}{f} \rangle \enspace . \]
  The auxiliary function $\cut_A : \Tri(E\times A) \to \Tri A$ is defined corecursively via
  \begin{align*} \comp{\cut}{\head} := \comp{\head}{\pr_2} \quad\text{ and } \quad
                     \comp{\cut}{\tail} := \comp{\tail}{\cut} \enspace . 
      \end{align*}
All the operations are suitably compatible with the bisimilarity relations, so that they can be equipped with the types
  \begin{align*}
    \redec_{A,B} &: \Setoid(\Tri A,\eq B) \to \Setoid(\Tri A,\Tri B ) \\
    \extend_{A,B} &: \Setoid(\Tri A,\eq B) \to \Setoid\bigr(\Tri (E \times A),\eq(E\times B)\bigr) \\
    \cut_A &:  \Setoid(\Tri(E\times A) , \Tri A) \enspace .
  \end{align*}
\end{example}

\begin{Long}
Note how heterogeneity of the destructor $\tail$ makes the definition of $\redec$ considerably more complicated than that of
the analogous operation $\sredec$ on streams.
\end{Long}

\section{Relative comonads and their morphisms}\label{sec:comonads}

In this section we define the category of \emph{comonads relative to a fixed functor}, and present some examples 
of such comonads and their morphisms.

\emph{Relative monads} were defined by \textcite{DBLP:conf/fossacs/AltenkirchCU10} as a notion of monad-like structure
whose underlying functor is not necessarily an endofunctor.
The dual notion is that of a relative \emph{co}monad:

\begin{definition}
 \label{def:rel_comonad}
  Let $F:\C\to\D$ be a functor. A \fat{relative comonad $T$ over $F$} is given by
  \begin{packitem}
   \item a map $T:\C_0 \to \D_0$ on the objects of the categories involved;
   \item an operation $\counit : \forall A : \C_0, \D(TA,FA)$;
   \item an operation $\cobind: \forall A,B:\C_0, \D(TA,FB) \to \D(TA,TB)$
   such that
   \item $\forall A,B:\C_0, \forall f:\D(TA,FB), \comp{\cobind(f)}{\counit_B} = f$;
   \item $\forall A : \C_0, \cobind(\counit_A) = \id_{TA}$;
   \item $\forall A,B,C:\C_0, \forall f : \D(TA,FB),\forall g:\D(TB,FC), \\
        \comp{\cobind(f)}{\cobind(g)} = \cobind(\comp{\cobind(f)}{g})$.
  \end{packitem} 
\end{definition}

\begin{Short}
\begin{remark}\label{def:lift}
 \begin{itemize}
  \item Relative comonads over the identity functor are exactly comonads.
  \item Just like relative monads, relative comonads are functorial.
  \item Under some conditions on the functor $F$, one can obtain comonads relative to $F$ from (absolute) comonads on the 
 source category $\C$.
  \item A \emph{weak constructive comonad} as defined by \textcite{DBLP:conf/types/MatthesP11}  is \emph{precisely}
  a comonad relative to the functor $\eq : \Set\to\Setoid$.
 \end{itemize}
\noindent
Details are given in the long version of this article \parencite{trimat_coq}.
\end{remark}
\end{Short}

\begin{Long}
Just like relative monads, relative comonads are functorial:
\begin{definition}
\label{def:lift}
 Let $T$ be a  comonad relative to $F:\C\to\D$.
 For $f : \C(A,B)$ we define
  $ \lift^T(f) := \cobind(\comp{\counit_A}{Ff}) : \D(TA,TB)$. 
 The functor properties are easily checked.
\end{definition}
\end{Long}

\begin{Long}
Relative comonads over the identity functor are exactly comonads.
\end{Long}

\begin{Long}

\begin{example}[Relative comonads from comonads]\label{ex_relcom_from_com}
  Let $F : \C \to \D$ be a fully faithful functor and $(M,\counit,\cobind)$ be a (traditional) comonad (in Kleisli form) on $\C$.
  We define a comonad $FM$ relative to $F$ by setting:
  \begin{itemize}
   \item $FM(A) := F(MA)$;
   \item $\counit^{FM}_A := F(\counit^M_A) : \D(FMA,FA)$;
   \item $\cobind^{FM}_{A,B} (f) := F\bigl(\cobind^M_{A,B}(F^{-1}f)\bigr)$.
  \end{itemize}
  The proof of the axioms of a relative comonad is immediate.
\end{example}

\end{Long}

\begin{example}[Streams]\label{ex_stream_comonad}
  The codata type family $\stream : \Set \to \Setoid$ of \Cref{ex_stream} is equipped with a structure of a comonad relative to the functor 
  $\eq : \Set \to \Setoid$ with
   $\counit_A := \shead_A$ and
   $\cobind_{A,B} := \sredec_{A,B}$.
\end{example}

\begin{Long}

\begin{example}[Trees]\label{ex_tree_comonad}
 Fix a type $B$. Analogously to \Cref{ex_stream_comonad}, the map $A \mapsto \Tree_B(A)$ of \Cref{ex_trees}
 is equipped with a structure of a comonad relative to $\eq: \Set\to\Setoid$.
\end{example}

\end{Long}

\begin{example}[Infinite triangular matrices]\label{ex:tri_comonad}
  The codata type family $\Tri : \Set \to \Setoid$ of \Cref{ex_tri} is equipped with a structure of a comonad relative to the functor 
  $\eq : \Set \to \Setoid$ with
   $\counit_A := \head_A$ and
   $\cobind_{A,B} := \redec_{A,B}$.
\end{example}

\begin{Long}
\begin{remark}
  A \emph{weak constructive comonad} as defined by \textcite{DBLP:conf/types/MatthesP11} to characterize the codata type $\Tri$
  and redecoration on it, is \emph{precisely}
  a comonad relative to the functor $\eq : \Set\to\Setoid$.
\end{remark}
\end{Long}

The notion of relative comonad captures many properties of $\stream$ resp.\ $\Tri$ and cosubstitution on them, in particular the interplay
of cosubstitution with the destructors $\shead$ resp.\ $\head$ via the first two axioms.
In order to  capture the interplay 
of cosubstitution  with the destructor $\stail$ resp.\ $\tail$,  we develop the notion of \emph{comodule over a relative comonad}
 in \Cref{sec:comodules}.

\begin{Long}
\emph{Morphisms of relative comonads} are natural transformations that are compatible with the comonadic structure:
\end{Long}
\begin{definition}
\label{def:comonad_morphism}
 Let $T$ and $S$ be comonads relative to a functor $F : \C \to \D$. A \fat{morphism of relative comonads} $\tau : T \to S$
  is given by a family of morphisms $\tau_A : \D(TA,SA)$ such that for any $A : \C_0$,
     $  \counit^T_A = \comp{\tau_A}{\counit^S_A} $
   and for any $A,B : \C_0$ and $f : \D(SA,FB)$,
   $  \comp{\cobind^T(\comp{\tau_A}{f})}{\tau_B} = \comp{\tau_A}{\cobind^S(f)}$.
\end{definition}

Relative comonads over a fixed functor $F$ and their morphisms form a category $\RComonad(F)$ with the obvious identity and composition operations.

\begin{remark}
A morphism $\tau : T\to S$ of relative comonads over a functor $F:\C\to\D$ is  \emph{natural}
with respect to the functorial action of \Cref{def:lift}.
\end{remark}

\begin{Long}

\begin{example}\label{ex_relcom_from_com_morphism}
 Continuing \Cref{ex_relcom_from_com} with $M, M'$ two monads on $\C$, given a comonad morphism $\tau : M \to M'$, one obtains a morphism of 
 relative comonads $F\tau : FM\to FM'$ by setting $F\tau_A := F(\tau_A)$.
 Again, the axioms are easy to check.
\end{example}

\begin{remark}
 The definitions given in \Cref{ex_relcom_from_com} and \Cref{ex_relcom_from_com_morphism} yield a functor from 
 comonads on $\C$ to comonads relative to $F:\C\to\D$. 
 If $F$ is a right adjoint with left adjoint $L$, $L\dashv F$, then postcomposing a comonad $T$ relative to $F$ with the functor $L$
 yields a monad on $\C$. Again, this map extends to morphisms.
 The two functors between categories of monads thus defined are again adjoints.
 Writing down the details is lengthy but easy.
 
 For instance, in a type theory with quotients, such as the \emph{Univalent Foundations} a.k.a.\ \emph{Homotopy Type Theory}
 \parencite{hottbook}, the functor \enquote{quotient} from setoids to types is left adjoint to the fully faithful 
 functor $\eq: \Set\to\Setoid$, thus above construction is applicable.
\end{remark}

\end{Long}

\begin{example}\label{ex_diag}
We define a morphism of relative comonads $\diagonal : \Tri \to \stream$:
Given a matrix $t : \Tri A$, its diagonal is a stream $\diagonal_A~t : \stream A$.
The map $\diagonal_A$ is defined via the clauses
\begin{align*} \comp{\diagonal_A}{\shead} := \head \quad\text{ and } \quad 
                  \comp{\diagonal_A}{\stail} := \comp{\comp{\tail}{\cut}}{\diagonal_A} \enspace . 
    \end{align*}
\end{example}

\begin{Long}

\begin{remark}
 The destructors $\stail$ (for $\stream$) and $\tail$ (for $\Tri$) are \emph{not} comonad morphisms.
 One can, however, equip the functor given by precomposing $\Tri$ with \enquote{product with $E$}, i.e.\
 $A \mapsto \Tri (E\times A)$, with a structure of relative comonad, induced by that
 on $\Tri$, cf.\ \Cref{product_comonad}.
\end{remark}

\begin{definition}\label{product_comonad}
  Let $T$ be a comonad relative to a strong monoidal functor $F:\C\to\D$ between cartesian monoidal categories,
  and let $E:\C_0$ be a fixed object of $\C$.
 The map $A\mapsto T(E\times A)$ inherits the structure of a comonad relative to $F$ from $T$: the 
 counit is defined as
   \[ \counit_A := \comp{\lift^T(\pr_2(E,A))}{\counit^T_{A}} \]
  and the cobind operation as
   \begin{align*} 
            \cobind_{A,B} : \D\bigl(T(E\times A),FB\bigr) &\to \D\bigl(T(E\times A),T(E\times B)\bigr) \\
              f &\mapsto  \cobind^T(\extend'~f)
   \end{align*}
  with $\extend'$ defined as 
  \begin{align*} \extend' : \D\bigl(T(E\times A),FB\bigr) &\to \D\bigl(T(E\times A), F(E\times B)\bigr) \enspace , \\ 
                                            f & \mapsto \comp{\langle \comp{T(\pr_1)}{\counit^T_E}, f \rangle}{{\phi^{F}_{E,B}}^{-1}} \enspace .
  \end{align*}
\end{definition}

\end{Long}

\section{Comodules over relative comonads}\label{sec:comodules}

In this section we develop the notion of \emph{comodule over a relative comonad}, dualizing the notion of module over a relative monad \parencite{ahrens_relmonads}.

\begin{definition}
\label{def:comodule}
 Let $T$ be a comonad relative to $F:\C\to\D$, and let $\E$ be a category.
 A \fat{comodule over T towards $\E$} consists of
   \begin{packitem}
   \item a map $M:\C_0 \to \E_0$ on the objects of the categories involved and
   \item an operation $\mcobind: \forall A,B:\C_0, \D(TA,FB) \to \E(MA,MB)$ such that
   \item $\forall A : \C_0, \mcobind(\counit_A) = \id_{MA}$;
   \item $\forall A,B,C:\C_0, \forall f : \D(TA,FB),\forall g:\D(TB,FC), \\
        \comp{\mcobind(f)}{\mcobind(g)} = \mcobind(\comp{\cobind(f)}{g})$ .
  \end{packitem}

\end{definition}

Every relative comonad comes with a canonical comodule over itself:

\begin{definition}
\label{def:tautological_comodule}
  Given a comonad $T$ relative to $F:\C\to\D$, the map $A \mapsto TA$ yields a comodule over $T$ 
  with target category $\D$, the \textbf{tautological comodule} of $T$, also called $T$.
  The comodule operation is given by
    $  \mcobind^T(f) := \cobind^T(f)$. 
\end{definition}

\begin{Long}
Similarly to relative comonads, comodules over these are functorial:
\begin{definition}
\label{def:comodule_lift}
 Let $M : \RComod(T,\E)$ be a comodule over $T$ towards some category $\E$. For $f : \C(A,B)$ we define
  \[ \mlift^M(f) := \mcobind(\comp{\counit_A}{Ff}) .  \]
\end{definition}
\end{Long}

A more interesting example of comodule is given by the functor that maps a type $A$ to the setoid $\Tri(E\times A)$
for some fixed type $E$:
\begin{example}\label{ex_tri_prod_comod}
   The map $A \mapsto \Tri(E\times A)$ is equipped with a comodule structure over the relative comonad $\Tri$ by
   defining the comodule operation $\mcobind$ as (cf.\ \Cref{ex_tri})
     $ \mcobind_{A,B} (f) := \redec (\extend~f)$.
\end{example}

A \emph{morphism of comodules} is given by a family of morphisms that is compatible with 
the comodule operation:

\begin{definition}
\label{def:morphism_of_comodules}
 Let $M, N : \C \to \E$ be comodules over the comonad $T$ relative to  $F:\C \to \D$.
 A \fat{morphism of comodules} from $M$ to $N$ is given by a family of morphisms 
   $ \alpha_A:\E(MA,NA) $
 such that for any $A,B:\C_0$ and $f : \D(TA,FB)$ one has
 $\comp{\mcobind^M(f)}{\alpha_B} = \comp{\alpha_A}{\mcobind^N(f)}$.
\end{definition}

 \begin{example}\label{ex_tail_comodule}
  The destructor $\stail_A : \stream A \to \stream A$ is the carrier of a morphism of tautological comodules (over the relative comonad $\stream$).
 \end{example}

\begin{example}\label{ex:tail_comodule}
 The destructor $\tail$ of \Cref{ex_tri} is a morphism of comodules over the comonad $\Tri$ 
  from the tautological comodule  $\Tri$ to the comodule $\Tri(E\times \_)$. 
\end{example}

Composition and identity of comodule morphisms happens pointwise. We thus obtain a category $\RComod(T,\E)$
 of comodules
over a fixed comonad $T$, towards a fixed target category $\E$.

\begin{Long}
\begin{remark}
  The family of morphisms constituting a comodule morphism is actually natural with respect to the functoriality 
  defined in \Cref{def:comodule_lift}.
\end{remark}
\end{Long}

Given a morphism of comonads, we can \enquote{transport} comodules over the source comonad to comodules over the target comonad:

\begin{definition}
\label{def:pushforward_comodule} 
  Let $\tau : T\to S$ be a morphism of comonads relative to a functor $F : \C \to \D$, and let furthermore $M$ be a 
  comodule over $T$ towards a category $\E$. We define the \fat{pushforward comodule} $\tau_*M$ to be the comodule over $S$ given by
  $  \tau_*M(A) := MA $
  and, for $f : \D(SA,FB)$,
   \[ \mcobind^{\tau_*M}(f) := \mcobind^M(\comp{\tau_A}{f}) : \E(MA,MB) \enspace . \]
   
  \noindent
  Pushforward is functorial: if $M$ and $N$ are comodules over $T$ with codomain category $\E$, and $\alpha : M\to N$ is 
    a morphism of comodules, then we define 
     $\tau_*\alpha : \tau_*M \to \tau_*N$
    as the family of morphisms
     $ (\tau_*\alpha)_A := \alpha_A$.
  It is easy to check that this is a morphism of comodules (over $S$) between $\tau_*M$ and $\tau_*N$.
  Pushforward thus yields a functor $\tau_*:\RComod(T,\E) \to \RComod(S,\E)$.
\end{definition}

As presented in \Cref{def:tautological_comodule}, every relative comonad induces a comodule over itself.
This extends to morphisms of relative comonads:

\begin{definition}
\label{def:induced} 
  Let $\tau : T\to S$ be a morphism of comonads relative to a functor $F : \C \to \D$.
  Then $\tau$ gives rise to a morphism of comodules over $S$ from the pushforward of the tautological comodule
  of $T$ along $\tau$ to the tautological comodule over $S$,
  \[ \induced{\tau} : \tau_*T \to S \enspace , \quad \induced{\tau}_A := \tau_A \enspace . \]
\end{definition}

\section{Terminality for streams and infinite triangular matrices}\label{sec:coalgebras_for_tri}

In this section, we define a notion of \enquote{coalgebra} for the signatures of streams and triangular matrices,
respectively. We then show that the codata types $\stream$ and $\Tri$ constitute the terminal object in
the respective category of coalgebras.
We put \enquote{coalgebra} in quotes for the reason that our coalgebras are not defined as coalgebras for a monad or an endofunctor.

The terminal coalgebra result is hardly surprising; however, it is still interesting as it characterizes not only the codata types themselves,
 but also the respective bisimilarity relations and comonadic operations on them, via a universal property.

\begin{Long}
\subsection{Coalgebras for $\stream$}

We first consider the homogeneous codata type of streams.
\end{Long}

\begin{definition}
 \label{cat_stream}
  A \fat{coalgebra for $\stream$} is given by a pair $(S,t)$ 
  consisting of
  \begin{itemize}
   \item a comonad $S$ relative to $\eq : \Set \to \Setoid$ and
   \item a morphism $t$ of tautological comodules over $S$, $t : S \to S$.
  \end{itemize}
  A coalgebra morphism $(S,t) \to (S',t')$ is given by a comonad morphism $\tau : S \to S'$ such that
     $ \comp{\tau_*t}{\induced{\tau}} = \comp{\induced{\tau}}{t'}$.
\end{definition}

This defines a category, with the obvious composition and identity. 

\begin{theorem}\label{thm_stream_terminal}
 The pair $(\stream, \stail)$ is the terminal coalgebra in the category of coalgebras of \Cref{cat_stream}.
\end{theorem}

More precisely, the aforementioned theorem says that the rules given in \Cref{stream_rules} allow to prove that
the category of coalgebras defined in \Cref{cat_stream} has a terminal object.

\begin{example}
  We equip the relative comonad $\Tri$ with the structure of a coalgebra for $\stream$ by defining a 
  morphism of tautological comodules over $\Tri$, given by
   $ t^{\diagonal} := \comp{\tail}{\cut}  : \Tri \to \Tri$.
  The resulting terminal coalgebra morphism
   $(\Tri, t^{\diagonal}) \to (\stream, \stail)$ has as underlying morphism of relative comonads the one defined in \Cref{ex_diag}.
\end{example}

\begin{Long}
 
\begin{remark}
 Fix a type $B$. A result analogous to \Cref{thm_stream_terminal} holds for trees $\Tree_B$ of \Cref{ex_tree_comonad}. 
 We refrain from giving a precise statement of this result.
\end{remark}

\end{Long}

\begin{Long}
\subsection{Coalgebras for $\Tri$}
\end{Long}

In analogy to the definition of coalgebras for the signature of streams, one would define
a coalgebra for the signature of $\Tri$ as a pair $(T,r)$ of a comonad $T$ relative to $\eq : \Set \to \Setoid$ and 
a morphism of comodules $r : T \to T(E\times \_)$. 
It turns out that in this way, one is not capable of obtaining the right auxiliary function $\cut$ for what is
supposed to be the \emph{terminal} such coalgebra (where $\cut$ is used to define the comodule $\Tri(E\times \_)$), namely the pair $(\Tri,\tail)$.
As a remedy, we define a coalgebra to come equipped with a specified operation analogous to $\cut$, and some laws governing
the behavior of that operation:

\begin{definition}
\label{def:rel_comonad_with_cut}
 Let $\C$ and $\D$ be categories with binary products and $F:\C\to\D$ a strong monoidal functor. Let $E:\C_0$ be a fixed object of $\C$.
 We define a \fat{comonad relative to $F$ with cut relative to $E$} to be a comonad $T$ relative to $F$ together with a $\cut$ operation 
    \[ \cut : \forall~A:\C_0, T(E\times A) \to TA \qquad \text{such that}\]
  \begin{itemize}
   \item $\forall~A:C_0, \comp{\cut_A}{\counit_A} = \comp{\lift^T(\pr_2(E,A))}{\counit_{A}}$;
   \item $\forall~A~B:C_0,\forall~f:\D(TA,FB), \comp{\cut_A}{\cobind(f)} = \comp{\cobind(\extend~f)}{\cut_B}$,
  \end{itemize}

  \noindent
  where, for $f:\D(TA,FB)$, we define $\extend(f) : \D\bigl(T(E\times A),F(E\times B)\bigr)$ as
       \[ \extend(f) := \comp{\comp{\langle T(\pr_1) , \cut \rangle}{(\counit_E\times f)}}{{\phi^{F}_{E,B}}^{-1}} \enspace . \]
  
\end{definition}

Morphisms of comonads with cut are morphisms of comonads that are compatible with the respective $\cut$ operations:

\begin{definition}
\label{def:morphism_comonad_cut}
 Let $(T,\cut^T)$ and $(S,\cut^S)$ be two comonads relative to a functor $F$ with cut relative to $E$ as in \Cref{def:rel_comonad_with_cut}.
 A \fat{morphism of comonads with cut} is a comonad morphism $\tau$ between the underlying comonads as in \Cref{def:comonad_morphism} that 
 commutes suitably with the respective $\cut$ operations, i.e.\ for any $A : \C_0$,
  $\comp{\tau_{E \times A}}{\cut^S_A}  = \comp{\cut^T_A}{\tau_A}$.
\end{definition}

Comonads with cut relative to a fixed functor $F:\C\to\D$ and $E:\C_0$ form a category $\RComonadWC(F,E)$.
There is the obvious forgetful functor from $\RComonadWC(F,E)$ to $\RComonad(F)$.
Conversely, any comonad $T$ relative to a suitable functor can be equipped with a $\cut$ operation, using functoriality of $T$.
\begin{Short}
 Details are given in the long version of this article.
\end{Short}

\begin{Long}
\begin{remark}[Canonical $\cut$ operation]\label{canonical_cut}
 Any comonad $T$ relative to a strong monoidal functor $F:\C\to\D$  can be equipped with a $\cut$ operation relative to 
 $E:\C_0$ satisfying the properties of \Cref{def:rel_comonad_with_cut} by setting
   \[ \ccut_A := \cut_A := \lift^T\bigl(\pr_2(E,A)\bigr) \enspace . \]
 (The extra \enquote{c} of $\ccut$ stands for \enquote{canonical}.)
 It follows from the axioms of comonad morphism that a comonad morphism $\tau : T\to S$ satisfies the equation of \Cref{def:morphism_comonad_cut} 
 for the thus defined operations $\ccut^T$ and $\ccut^S$, hence constitutes a morphism of comonads with cut from $(T,\ccut^T)$ to $(S,\ccut^S)$.
 We thus obtain a functor 
 \[ \ccut_{F,E} : \RComonad(F) \to \RComonadWC(F,E)\]
 from relative comonads over $F$ to relative comonads over $F$ with cut relative to a fixed object $E:\C_0$ given on 
 objects by $T\mapsto (T,\ccut^T)$.
\end{remark}

The functor $\ccut_{F,E}$, followed by the forgetful functor, yields the identity. We can thus view
relative comonads with cut as a generalization of relative comonads.
\end{Long}

\begin{Long}
Our prime example of relative comonad comes with a $\cut$ operation that is \emph{not} the canonical one:
\end{Long}

\begin{example}
\label{def:cut_for_tri}
  The relative comonad $\Tri$ from \Cref{ex:tri_comonad}, together with the $\cut$ operation defined in \Cref{ex_tri}, 
  is a comonad with cut as in \Cref{def:rel_comonad_with_cut}.
\end{example}

Given a comodule $M$ over a relative comonad $T$ with cut, we define a comodule over $T$ obtained by precomposition of $M$ with
\enquote{product with a fixed object $E$}:

\begin{definition}
\label{def:product_in_context}
 Suppose $F:\C\to\D$ is a strong monoidal functor, and $T$ is a comonad relative to $F$ with a $\cut$ operation 
 relative to $E:\C_0$ as in \Cref{def:rel_comonad_with_cut}.
 Given a comodule $M$ over $T$,  \fat{precomposition with} \enquote{\fat{product} with $E$}
 gives a comodule $M(E\times\_) : A \mapsto M(E\times A) $ over $T$.
 The comodule operation is deduced from that of $M$ by 
 \begin{align*} \mcobind^{M(E\times\_)}_{A,B} : \D(TA,FB)&\to \E\bigl(M(E\times A), M(E\times B)\bigr) \enspace ,\\ 
                                                      f &\mapsto \mcobind^M_{E\times A,E\times B}(\extend(f)) \enspace ,
  \end{align*}                                        
where the $\extend$ operation is the one defined in \Cref{def:rel_comonad_with_cut}.
 
 Furthermore, given two comodules $M$ and $N$ over $\T$ with target category $\E$, and a comodule morphism $\alpha : M \to N$,  
 the assignment $ \alpha(E \times \_)_A := \alpha_{E\times A}$ defines a comodule morphism 
  $\alpha(E\times \_) : M(E\times \_) \to N(E\times \_) $.

\begin{Long}
  \noindent
  We thus obtain an endofunctor on the category of comodules over $T$ towards $\E$,
   $ M \mapsto  M (E\times \_) : \RComod(T,\E) \to \RComod(T,\E)$.
\end{Long}
\end{definition}

\begin{remark}[Pushforward commutes with product in context]\label{rem:prod_pullback_commute}
 Note that the constructions of \Cref{def:product_in_context} and \Cref{def:pushforward_comodule} commute:
 we have an isomorphism of comodules 
  $\tau_*(M(E\times \_)) \cong (\tau_*M)(E \times \_)$
 given pointwise by identity morphisms.
\end{remark}

\begin{Long}

It directly follows from the definition that the cut operation of any comonad $T$ with cut 
constitutes a comodule morphism $\cut : T(E\times \_) \to T$.
We can thus restate the definition of a morphism of comonads with cut as in \Cref{def:morphism_comonad_cut} by asking the following diagram 
of comodule morphisms (in the category $\RComod(S,\D)$) to commute
(where in the upper left corner we silently add an isomorphism as in \Cref{rem:prod_pullback_commute}):
 \[ \begin{xy}
       \xymatrix{  **[l] \tau_*T(E\times \_ )  \ar[r]^{\tau_*(\cut^T)} \ar[d]_{\induced{\tau}(E\times \_)}  &  **[r] \tau_*T \ar[d]^{\induced{\tau}} \\
                   **[l]  S (E\times \_ ) \ar[r]_{\cut^S}  &  **[r] S  \enspace .
        }
      \end{xy}
   \]

\end{Long}

\begin{Long}
The construction of \Cref{def:product_in_context} yields a categorical characterization of the $\tail$ destructor---%
more precisely, of its behavior with respect to cosubstitution as in \Cref{eq:rest_redec}---via the notion of comodule morphism:

\begin{example}\label{ex:tail_comodule_alternative}
This example is a reformulation of \Cref{ex:tail_comodule}.
 Consider the comonad $\Tri$, equipped with the $\cut$ operation of \Cref{def:cut_for_tri}.
 The destructor $\tail$ of \Cref{ex_tri} is a morphism of comodules over the comonad $\Tri$ 
  from the tautological comodule  $\Tri$ to $\Tri(E\times \_)$.
  
\end{example}
\end{Long}

\begin{definition}
\label{def:cat_tri}
   Let $E:\Set_0$ be a set.
   Let $\mathcal{T} = \mathcal{T}_E$ be the category of \fat{coalgebras for infinite triangular matrices} where an object consists of
   \begin{itemize}
    \item a comonad $T$ over the functor $\eq:\Set\to\Setoid$ with $\cut$ relative to $E$ and
    \item a morphism $\tail$ of comodules over $T$ of type $T \to T(E\times \_)$
   \end{itemize}
   such that for any set $A$,
    $ \comp{\cut_A}{\tail_A} = \comp{\tail_{E\times A}}{\cut_{E\times A}}$.
    
 \begin{Long}
   The last equation can be stated as an equality of comodule morphisms as
     \[ \comp{\cut}{\tail} = \comp{\tail(E\times \_)}{\cut(E\times\_)} \quad \bigl( = (\comp{\tail}{\cut})(E\times \_)\bigr)\enspace . \]
\end{Long}

   A morphism between two such objects $(T,\tail^T)$ and $(S,\tail^S)$
   is given by a morphism of relative comonads with cut $\tau : T \to S$ such that
   the following diagram of comodule morphisms in the category $\RComod(S,\E)$ commutes,
   
   \[ \begin{xy}
       \xymatrix{   \tau_*T  \ar[r]^{\tau_*(\tail^T)} \ar[d]_{\induced{\tau}}  &  **[r] \tau_*T (E\times \_ )\ar[d]^{\induced{\tau}(E\times \_)} \\
                    S  \ar[r]_{\tail^S}  &  **[r] S (E\times \_ ) \enspace .
        }
      \end{xy}
   \]

   \noindent
   Here in the upper right corner we silently insert an isomorphism as in \Cref{rem:prod_pullback_commute}.
\end{definition}

\begin{theorem}\label{ex:final_sem_tri} 
   The pair $(\Tri, \tail)$ consisting of the relative comonad with cut $\Tri$ of \Cref{def:cut_for_tri} together with 
    the morphism of comodules $\tail$ of \Cref{ex:tail_comodule},
   constitutes the terminal coalgebra of triangular matrices.
\end{theorem}

\begin{Long}
\begin{proof}[sketch]
   For a given coalgebra $(T,\tail^T)$, the (terminal) morphism $\bigcirc = \bigcirc_T:T\to\Tri$ is defined via the corecursive equations
    \begin{align}\head\bigl(\bigcirc~t\bigr) &:= \counit^T~t \quad\text{ and } \label{eq:terminal_top}\\
                     \tail\bigl(\bigcirc~t\bigr) &:= \bigcirc(\tail^T~t) \enspace . \label{eq:terminal_rest}
      \end{align}
      By coinduction we show that the map $\bigcirc$ is compatible with $\cobind$ and $\cut$ operations of the source and 
   target coalgebras. We omit these calculations, which can be consulted in the \coq source files.
   
   Note that there is actually no choice in this definition: \Cref{eq:terminal_top} is forced upon us since we want $\bigcirc$ to constitute 
   a morphism of comonads---the equation directly corresponds to one of the axioms.
   \Cref{eq:terminal_rest} is forced upon us by the diagram a morphism of coalgebras has to make commute.
   
   The same argument is used to show, again by coinduction, that any two morphisms of coalgebras $\tau,\rho : (T,\tail^T) \to (\Tri,\tail)$
   are equal, thus concluding the proof.   
\end{proof}
\end{Long}

This universal property of terminality characterizes not only the codata type of infinite triangular matrices, but also
the \fat{bisimilarity} relation on it as well as the \fat{redecoration} operation.

\begin{Long}
\section{Formalization in \coq}\label{sec:formal}

All our definitions and theorems are mechanized in the proof assistant \coq \parencite{coq84pl3}.
The formalization of infinite triangular matrices is taken from the work by \textcite{DBLP:conf/types/MatthesP11},
and only slightly adapted to compile with the version of \coq we use.
The mechanization does \emph{not} rely on any additional axioms.
The \coq source files and HTML documentation are available from the project web site \parencite{trimat_coq}.

In the following we explain some of our design choices for this mechanization
and point out differences between the pen-and-paper definitions and the mechanized ones.

\subsection{Implementation choices}

We explain two choices we made in the course of the formalization in \coq. The first choice concerns
the formalization of categories, more precisely, how to formalize \emph{equality of morphisms}.
The second choice concerns the formalization of algebraic structures.

\subsubsection{Setoids for hom-sets}
We formalize categories to be given by a type of objects and a dependent type---indexed by pairs of objects---of morphisms,
equipped with suitable composition and identity operations satisfying appropriate axioms.
More precisely, the family of morphisms is given by a family of \emph{setoids}, where the setoidal equivalence relation on each
type of morphisms denotes the equality relation on these morphisms. This approach was first used by
\textcite{aczel_galois} in the proof assistant \texttt{LEGO}, and also by \textcite{concat}  in their library
of category theory in \coq.
At the moment, it seems to be the standard way of formalizing categories in intensional Martin-L\"of type theory.
Alternatively, we could have chosen to consider morphisms modulo propositional equality, which is feasible in a more extensional type theory \parencite{rezk_completion}.

Indeed, the morphisms we consider---morphisms of comonads and comodules---are given by structures
bundling a lot of data and properties; in order to consider two such morphisms as equal, we usually only compare one field of the 
corresponding records. Furthermore, this field usually consists of a (dependent) function.
It would be rather cumbersome to reduce equality of two such records to extensional equality of one of their fields, 
necessitating the use of the axioms of propositional and functional extensionality in IMLTT.
Using setoids for morphisms instead seems to come with less overhead and to be conceptually cleaner.

\subsubsection{Records vs.\ classes}
Two approaches to the formalization of mathematical structures have been used extensively in \coq: on the one hand, packaging structures
in \emph{record types}  in combination with use of \emph{canonical structures}, is used with success, e.g., in 
the formalization of algebraic structure in the context of the proof of the Feit-Thompson theorem \parencite{DBLP:conf/tphol/GarillotGMR09}.
On the other hand, \textcite{DBLP:journals/mscs/SpittersW11} suggest the use of \emph{type classes}, in particular when multiple inheritance
is an issue.

In the present formalization, we decide to use records rather than classes, since the strongest argument for type classes---multiple inheritance---does 
not occur.
We make use of canonical structures in order for \coq to deduce instances of categories when we mention objects of a category; 
in particular, this is used to allow for overloading of the notation for morphisms of a category.
We can thus conveniently use the same arrow symbol to denote the type of morphisms between two comonads, between two comodules and so on.

\subsection{Formal vs.\ informal definitions}

In the \coq formalization, we provide two different versions of the terminal semantics results:

In one version, we use the \lstinline!CoInductive! vernacular command of \coq to define the codata types we consider.
In this way, coinductive types are specified through \fat{constructors} rather than \fat{destructors}. The definition
of functions into coinductive data types thus specified hence looks very different to function definitions 
in terms of destructors as used in the present article.

In the other version we do not use that device for specifying coinductive types in \coq, but augment \coq by the axioms given in \Cref{stream_rules} and \Cref{tri_rules},
respectively, and prove the existence of a terminal coalgebra from these axioms.
This version thus is in close correspondence with the theory presented in the article.

This might be the right moment to point to work on a device allowing the declaration of coinductive types via destructors in \agda, 
see \parencite{DBLP:conf/popl/AbelPTS13}.

\end{Long}

\section{Conclusions and future work}

We have given a category-theoretic characterization, via a universal property, of streams and of infinite triangular matrices,
each equipped with a cosubstitution operation,
in intensional Martin-L\"of type theory.

The development of a notion of \enquote{signature} at least for \emph{homogeneous} codata types, and 
a terminal semantics for them, in line with the definitions we give for $\stream$, is easy
and will be treated in a forthcoming work.

In a more extensional type theory, such as \emph{Homotopy Type Theory} \parencite{hottbook}, 
one can reflect bisimilarity into propositional equality by \emph{quotienting},
thus eliminating the need to work with setoids.
This will be investigated in future work.

Furthermore, we will work on a suitable notion of \emph{signature} for the specification of general coinductive data types with
a cosubstitution operation.

Finally, we would like to integrate \emph{equations} into the notion of signature, which 
will allow, e.g., considering branching trees modulo permutation of subtrees.

\begin{Long}
 
\subsubsection*{Acknowledgments}
 We thank André Hirschowitz, Ralph Matthes and Paige North for many helpful discussions.

\end{Long}
 

\bibliographystyle{plain}
\bibliography{literature}

\appendix

\input{stream_rules}

\input{tri_rules}

\input{formal_table}

\end{document}

%% file: stream_rules.tex
\section{Rules for $\stream$ and bisimilarity}\label{stream_rules}

\subsection{$\stream$}

\begin{description}

 \item[Formation]\hfill \\
 
 \begin{center}
 \def\extraVskip{3pt}
     \def\proofSkipAmount{\vskip.8ex plus.8ex minus.4ex}

   \AxiomC{$A : \Set$}
    \UnaryInfC{$\stream A : \Set$}
     \DisplayProof
 \end{center} 
 
 \item[Destruction]\hfill \\


\begin{center}
    \AxiomC{$t : \stream A$} 
     \UnaryInfC{$\shead_A~t : A$}
      \DisplayProof
                        \hspace{3ex}
                                       \AxiomC{$t : \stream A$}
                                       \UnaryInfC{$\stail_A~t : \stream A$}
                                       \DisplayProof%
\end{center}
  \item[Creation]\hfill \\                                     
                       
            
\begin{center}
               \AxiomC{$T : \Set$} \AxiomC{$hd : T \to A$} \AxiomC{$tl : T \to T$} 
               \TrinaryInfC{$\corec_A~hd~tl : T \to \stream A$}
               \DisplayProof%
\end{center}
                      
  \item[Computation]\hfill \\

\begin{center}          
               \AxiomC{$hd : T \to A$} \AxiomC{$tl : T \to T$} \AxiomC{$t:T$}
               \TrinaryInfC{$\shead_A(\corec_A~hd~tl~t) = hd(t)$}
               \DisplayProof
               
               \vspace{1em}
               
               \AxiomC{$hd : T \to A$} \AxiomC{$tl : T \to T$} \AxiomC{$t:T$}
               \TrinaryInfC{$\stail_A(\corec_A~hd~tl~t) = \corec_A~hd~tl~(tl~t)$}
               \DisplayProof
\end{center}

 \end{description}              
               
\subsection{Bisimilarity}\label{stream_bisim}

\begin{description}

 \item[Formation]\hfill \\
 
 \begin{center}
 \def\extraVskip{3pt}
     \def\proofSkipAmount{\vskip.8ex plus.8ex minus.4ex}

   \AxiomC{$A : \Set$} \AxiomC{$s, t : \stream A$}
    \BinaryInfC{$\bisim_A~s~t : \Set$}
     \DisplayProof
 \end{center} 
 
 \item[Destruction]\hfill \\


\begin{center}
    \AxiomC{$s, t : \stream A$} \AxiomC{$p: \bisim_A~s~t$} 
     \BinaryInfC{$\shead_A~s = \shead_A~t$}
      \DisplayProof
                        \hspace{3ex}
                                       \AxiomC{$s,t : \stream A$} \AxiomC{$p: \bisim_A~s~t$}
                                       \BinaryInfC{$\bisim_A (\stail_A~s) (\stail_A~t)$}
                                       \DisplayProof%
\end{center}
  \item[Creation]\hfill \\                                     
                       
            
\begin{center}
               \AxiomC{$R : \stream A \to \stream A \to \Set$} \noLine\UnaryInfC{$\forall~s,t : \stream A, R~s~t \to \shead~s = \shead~t$} \noLine
                \UnaryInfC{$\forall~s,t : \stream A, R~s~t \to \bisim (\stail~s) (\stail~t)$}  
               \UnaryInfC{$\forall~s,t : \stream A, R~s~t \to \bisim~s~t$}
               \DisplayProof%
\end{center}
                      

 \end{description}

%% file: tri_rules.tex
\section{Rules for $\Tri$ and bisimilarity}\label{tri_rules}

\subsection{$\Tri$}

\begin{description}

 \item[Formation]\hfill \\
 
 \begin{center}
 \def\extraVskip{3pt}
     \def\proofSkipAmount{\vskip.8ex plus.8ex minus.4ex}

   \AxiomC{$A : \Set$}
    \UnaryInfC{$\Tri A : \Set$}
     \DisplayProof
 \end{center} 
 
 \item[Destruction]\hfill \\


\begin{center}
    \AxiomC{$t : \Tri A$} 
     \UnaryInfC{$\head_A~t : A$}
      \DisplayProof
                        \hspace{3ex}
                                       \AxiomC{$t : \Tri A$}
                                       \UnaryInfC{$\tail_A~t : \Tri(E\times A)$}
                                       \DisplayProof%
\end{center}
  \item[Creation]\hfill \\                                     
                       
            
\begin{center}
               \AxiomC{$T : \Set\to\Set$} 
               \AxiomC{$hd : \forall A,TA \to A$} \AxiomC{$tl : \forall A, T A \to T(E\times A)$} 
               \TrinaryInfC{$\corec_T~hd~tl :  \forall A, T A \to \Tri A$}
               \DisplayProof%
\end{center}
                      
  \item[Computation]\hfill \\

\begin{center}          
               \AxiomC{$hd : \forall A,TA \to A$} \AxiomC{$tl : \forall A, T A \to T(E\times A)$} 
               \AxiomC{$t : TA$}
               \TrinaryInfC{$\head_T(\corec_A~hd~tl~t) = hd(t)$}
               \DisplayProof
               
               \vspace{1em}
               
               \AxiomC{$hd : \forall A,TA \to A$} \AxiomC{$tl : \forall A, T A \to T(E\times A)$} 
               \AxiomC{$t : TA$}
               \TrinaryInfC{$\tail_T(\corec_A~hd~tl~t) = \corec_A~hd~tl~(tl~t)$}
               \DisplayProof
\end{center}

 \end{description}              
               
\subsection{Bisimilarity}

 \begin{description}

 \item[Formation]\hfill \\
 
 \begin{center}
 \def\extraVskip{3pt}
     \def\proofSkipAmount{\vskip.8ex plus.8ex minus.4ex}

   \AxiomC{$A : \Set$} \AxiomC{$s, t : \Tri A$}
    \BinaryInfC{$\bisim_A~s~t : \Set$}
     \DisplayProof
 \end{center} 
 
 \item[Destruction]\hfill \\


\begin{center}
    \AxiomC{$s, t : \Tri A$} \AxiomC{$p: \bisim_A~s~t$} 
     \BinaryInfC{$\head_A~s = \head_A~t$}
      \DisplayProof
                        \hspace{3ex}
                                       \AxiomC{$s,t : \Tri A$} \AxiomC{$p: \bisim_A~s~t$}
                                       \BinaryInfC{$\bisim_A (\tail_A~s) (\tail_A~t)$}
                                       \DisplayProof%
\end{center}
  \item[Creation]\hfill \\                                     
                       
            
\begin{center}
               \AxiomC{$R : \forall A, \Tri A \to \Tri A \to \Set$} \noLine
               \UnaryInfC{$\forall A,\forall~s,t : \Tri A, R~s~t \to \head~s = \head~t$} \noLine
                \UnaryInfC{$\forall  A,\forall~s,t : \Tri A, R~s~t \to \bisim (\stail~s) (\stail~t)$}  
               \UnaryInfC{$\forall A,\forall~s,t : \Tri A, R~s~t \to \bisim~s~t$}
               \DisplayProof%
\end{center}

\end{description}

%% file: formal_table.tex
\section{Correspondence of informal and formal definitions}\label{sec:table_formal_informal}

All our definitions and theorems are formalized in the proof assistant \coq.
The \coq files and HTML documentation are available from the project web page \parencite{trimat_coq}.
For easier orientation, the table below gives the correspondence between the items in this article and
their names in the formal development.

{

\Crefname{definition}{Def.}{Defs.}
\Crefname{theorem}{Thm.}{Thms.}
\Crefname{example}{Ex.}{Exs.}

\begin{center}
{\renewcommand{\arraystretch}{1.2}
\begin{tabular}{lllll}
Informal && Reference && Formal \\ \hline
Category &&  && \lstinline!Category!\\
Functor &&  && \lstinline!Functor!\\
Relative comonad && \Cref{def:rel_comonad} && \lstinline!RelativeComonad!\\
Triangular matrices as comonad && \Cref{ex:tri_comonad} && \lstinline!Tri!\\
Comodule over comonad && \Cref{def:comodule} && \lstinline!Comodule!\\
Tautological comodule (of $T$) && \Cref{def:tautological_comodule} &&\lstinline!tcomod!, \lstinline!<T>!\\
$\stail$ is comodule morphism &&\Cref{ex_tail_comodule}&& \lstinline!Tail!\\
$\tail$ is comodule morphism &&\Cref{ex:tail_comodule} && \lstinline!Rest!\\
Pushforward comodule && \Cref{def:pushforward_comodule} && \lstinline!pushforward!\\
Induced comodule morphism &&\Cref{def:induced} && \lstinline!induced_morphism!\\
Coalgebras for streams    &&\Cref{cat_stream} && \lstinline!Stream!\\
$\stream$ is terminal && \Cref{thm_stream_terminal} && \lstinline!StreamTerminal.Terminality!\\
Relative comonad with cut &&\Cref{def:rel_comonad_with_cut} && \lstinline!RelativeComonadWithCut!\\
Precomposition with product && \Cref{def:product_in_context} &&\lstinline!precomposition_with_product!\\
Coalgebras for triangular matrices && \Cref{def:cat_tri} && \lstinline!TriMat!\\
$\Tri$ is terminal && \Cref{ex:final_sem_tri} && \lstinline!TriMatTerminal.Terminality!\\
\end{tabular}
}
\end{center}

}